# THE HEALTH STATUS OF A POPULATION: HEALTH STATE AND SURVIVAL CURVES, AND HALE ESTIMATES*

**Christos H Skiadas**[1] **and Charilaos Skiadas**[2]

[1]ManLab, Technical University of Crete, skiadas@cmsim.net

[2]Department of Mathematics and Computer Science, Hanover College, Indiana, USA, skiadas@hanover.edu

---

**Abstract**

In this paper we explore the very important case of finding a health measure in the lines of the survival curve but independent of the standard deviation parameter $\sigma$. This is done by estimating the health state curve and calculating the total area under the curve. An interesting comparison with the survival curves is done. The health status measure (HSM or HS) resulting is of the form of the life expectancy (LE) as it is expressed in terms of years of age. The HS is independent of the standard deviation $\sigma$. When a perfect rectangularization of the survival curve appear, the LE is equal to the age where the health state curve approaches zero. The provided HS form gives rise to an interesting classification of various countries. The results are close to those obtained by the LE estimator with an interesting reorganization of country ranking. We also provide illustrations of our estimated Health Status along with comparative presentations of the Healthy Life Expectancy and the HALE measures in several countries. Theoretical issues are provided and important stochastic simulations are done along with reproduction of the death probability density forms.

## 1 Introduction

The form of the survival curve resulting from a classical life table is a measure of the health state or health status of a population. Much

---

attention has given to the exploration of the part of the influence of disability on the form of the survival curve thus providing methods for estimating the loss of healthy life years in a population. However, the form of the survival curve depends on the dispersion of deaths around a mean value, measured by the standard deviation $\sigma$ thus making the survival curve not very appropriate when doing comparisons between different populations.

An approach based on a simple Inverse Gaussian model and applied to Carey Medfly data is due to Weitz and Fraser (2001). The Inverse Gaussian form cannot apply to the human population data. The needed general form is given in Skiadas and Skiadas (2010) whereas the theoretical and technical issues are presented in Janssen and Skiadas (1995).

The main points in estimating the HSM are: to introduce a "system of measure" a "metric" that is independent of the standard deviation $\sigma$, to effectively estimate the health state, to reproduce the population behavior, to compare the health state in various countries and populations for the same or various time periods, to be able to calculate to cost of the health improvement of the population, to estimate the insurance cost and the insurance policies.

## 1.1 The Related Theory

Following the first exit time theory we assume that the health state of an individual is expressed by a stochastic function denoted by $S_x$ and the associated stochastic paths over time $t$ or age $x$ are estimated after integrating the stochastic differential equation

$$dS_x = h_x dx + \sigma \, dW_x \qquad (1)$$

With drift $h_x$ and finding the formula for the stochastic paths $S_x$

$$S_x = \int_0^x h_s ds + \int_0^x dW_s = H_x + \sigma W_x \qquad (2)$$

Where $W_x$ is the Wiener process and the Health State $H_x$ is provided as the integral of the instantaneous change $h_x$

$$h_x = \frac{dH_x}{dx} \tag{3}$$

The death occurs when $S_x = 0$ and from (2) follows that

$$H_x + \sigma W_x = 0 \tag{4}$$

The simpler form for the Health State $H_x$ should be a decreasing process of the form (see related bibliography in Janssen and Skiadas 1995 and Skiadas 2010, 2013, 2014, 2015):

$$H_x = l - (bx)^c \tag{5}$$

Where *l*, *b and c* are parameters. The form of (4) becomes

$$l - (bx)^c + \sigma W_x = 0 \tag{6}$$

This is a very important relation providing different ways of simulation of the stochastic process.

This is demonstrated by observing the new form of (2) that is

$$S_x = l - (bx)^c + \sigma W_x \tag{7a}$$

This form is important for constructing stochastic paths for the health state $S_x$.

The stochastic paths (*l*=0) are presented in the next Figure 1A where the Health State curve is expressed by the heavy red curve, the confidence intervals by the dashed blue curves and the stochastic paths are illustrated by the light lines. Every stochastic path expresses the health state of an individual during the life course. The end comes when the stochastic path reaches the level zero in a particular age. Then a point is added to the previous points produced in this age and in this particular place. The same is done for all the 117 age years studied producing the histogram for the death probability density of Figure 1B after 260.000 stochastic realizations. The data are from the life table of the Human Mortality Database (HMD) for USA females in 2010. The produced simulations are very close to the real data as is expressed by observing the histograms (note that as the data in the life tables are constructed for 100000 population and the total area of the death probability density should be equal to one we have to divide the numbers for D(x) provided by the HMD by 100000). As we have done successive simulations we can easily find the Life Expectancy from the simulations to be 81.41 years of age very close to the 81.22 years provided by the HMD. Note that in our simulations we have not taken into consideration the infant mortality.

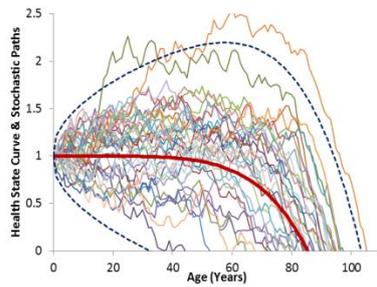

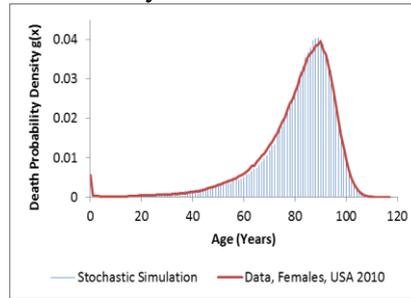

Fig. 1A. Health State Curve and Stochastic paths

Fig. 1B. Death Probability Density

The next form arises from (7a) by a simple transformation

$$S_x + (bx)^c = l + \sigma W_x \qquad (7b)$$

In this case we obtain the same simulations' results for the death probability density (see Figure 2B) as from the previous case though the simulation process presented in Figure 2A is different. In this case the stochastic paths generated from the standard Wiener process start from $l=1$ and are developed horizontally until to reach for the first time the curve expressed by the curve $z(x)=(bx)^c$. This case is more appropriate from an explanatory point of view as the curve expressed by $zf(x)=(bx)^c$ follows a continuously increasing path analogous to the force of mortality curve. The application is for the USA females at in 2010.

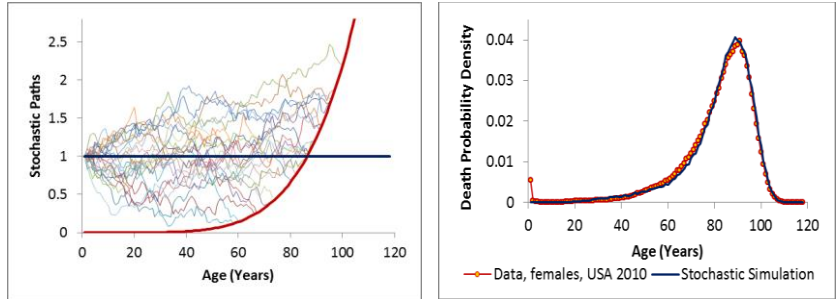

Fig. 2A. Health State Curve and Stochastic paths

Fig. 2B. Death Probability Density

A third form is given by

$$S_x - l + (bx)^c = \sigma W_x \qquad (7c)$$

The three forms (7a), (7b) and (7c) are mathematically the same. However, they provide three distinct simulation opportunities very important to explain the development of the health status and the development of the death probability density function.

According to the theory developed (Janssen and Skiadas, 1995) first it was solved the associated Fokker-Planck equation for the appropriate boundary conditions in order to find the transition probability density function.

In the general case we have worked with the form

$$dS_x = h_x dx + \sigma_x dW_x$$

obtaining the following transition probability density form

$$f(x) = \frac{1}{\int_0^x \sigma(s)ds \sqrt{2\pi x}} e^{-\frac{H_x^2}{2x\left(\int_0^x \sigma(s)ds\right)^2}}$$

Where $\qquad H_x = \int_0^x h_s ds$

For a stable $\sigma$ the simple form arises

$$f(x) = \frac{1}{\sigma\sqrt{2\pi x}} e^{-\frac{H_x^2}{2\sigma^2 x}}$$

Then we could find the formula for the first exit time probability density function $g(x)$ for the health state stochastic process crossing or hitting for the first time a barrier set to zero.

$$g(x) = \frac{\left| H_x - t H_x' \right|}{\sigma\sqrt{2\pi x^3}} e^{-\frac{(H_x)^2}{2\sigma^2 x}} \tag{8}$$

## 2   The Weitz and Fraser Paper Revisited

Fifteen years after the Weitz and Fraser (2001) paper only few applications appear though the high importance of this paper for the establishment of a system of measuring the health state or health status of a population. Perhaps the few applications are due to the paper title "Explaining mortality rate plateaus" instead of the use of the first exit time or hitting time theory in relation to the health status.

Weitz and Fraser introduced the simpler form for the health state, presented here by equation (5), but with the parameter $c = 1$. This is a form not convenient for the human population health status but it could be used to model the health status of a large group of Medflies systematically studied by Carey (1992).

$$H_x = l - (bx) \tag{9}$$

This form for the health status provides the following probability density function for the first exit time from a barrier set at zero level, that is:

$$g(x) = \frac{l}{\sigma\sqrt{2\pi x^3}} e^{-\frac{(l-bx)^2}{2\sigma^2 x}} \tag{10}$$

This is the classical form of the so-called Inverse Gaussian known at least from 1915.

When trying to estimate the parameters of (10) from the mortality data $d(x)$ or $g(x)$ by a non-linear regression analysis it is obvious that we cannot estimate simultaneously $\sigma$ with the two parameters $b$ and $l$. Instead by using the transformation $b*=b/\sigma$ and $l*=l/\sigma$ we arrive in the following (11) form providing the new parameters $b*$ and $l*$ as functions of $\sigma$. Instead Weitz and Fraser selected to set $l=1$ so that they estimated $\sigma$ and $b$. Both methods can be useful for the estimates that follow. Note that setting a health state level equal to unity was proposed by Torrance (1976).

$$g(x) = \frac{(l/\sigma)}{\sqrt{2\pi x^3}} e^{-\frac{((l/\sigma)-(b/\sigma)x)^2}{2x}} = \frac{l*}{\sqrt{2\pi x^3}} e^{-\frac{(l*-b*x)^2}{2x}} \qquad (11)$$

## 2.1 Deterministic and Stochastic Case

An interesting question is related to the deterministic and stochastic case of the problem by means of estimating the mean value of the process. This is illustrated in Figure 3A where the mean value of the deterministic process is the line AB and the mean value of the stochastic process is precisely the same line AB. That is different is that the deterministic line is the result of the mean value of $N$ members of the population identically distributed with $\sigma=0$ whereas, in the stochastic case, the mean value provided in line AB is the average of the stochastic paths of $N$ members of the population. Although no-stochastic path is identical to another the mean value is identical to the deterministic case. That it is different is that $\sigma>0$. From (2) and (9) the stochastic paths are of the form:

$$S_x = H_x + \sigma W_x = l - bx + \sigma W_x \qquad (12)$$

The deterministic case ($\sigma=0$) is given by:

$$S_x = H_x = l - bx \qquad (13)$$

Following the deterministic case death occurs at $H_x=0$ at age $x=l/b$ (point B of Figure 3A). For the stochastic case the estimation procedure based on (11) will estimate $l*=l/\sigma$ and $b*=b/\sigma$. For this case death occurs at $H_x=0$ at age $x= l^*/ b^*=( l/\sigma)/( b/\sigma)=l/b$ that is precisely the same as in the deterministic case as it is independent of $\sigma$.

Note that the stochastic case provides the death probability density form of Figure 3B whereas the deterministic case will distribute all deaths exactly at point B of Figure 3A thus producing a sharp distribution form at Figure 3B. The higher the standard deviation $\sigma$ results in larger dispersion in the death distribution.

Instead the starting point of the process (the health state $l$ at birth) is at level $l$ for the deterministic process (point A, Figure 3A) whereas it is at level $l^*=l/\sigma$ for the stochastic process by means that the starting point for the stochastic process resulting from the estimation from (11) can be in a large variety of places in the $Y$ axis depending on the selection of $\sigma$ by means that the location of the point A is not definite and the same holds for the estimation of the level of the health state at birth. Instead it is perfectly located the zero health state at point B (Figure 3A).

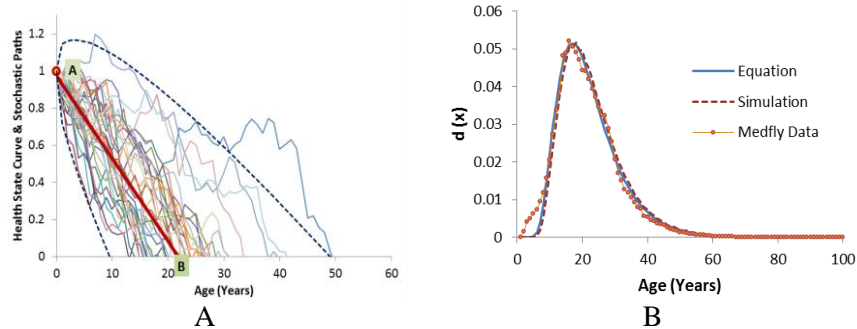

Fig. 3. A. Stochastic paths and B. Stochastic simulation for Carey Medfly

For the deterministic case we can estimate the total health state estimated as the area of the triangle (OAB) that is $H_{total}=(OA)*(OB)/2=l.(l/b)/2=(l^2/b)/2$. The mean value is $H_{mean}=l/2$

For the stochastic case the total health state is $H_{total}=(OA)*(OB)/2=l^*.(l/b)/2=(l/\sigma)(l/b)/2$. The mean value is $H_{mean}=l*/2=(l/\sigma)/2$. This value is equal to the deterministic case only when $\sigma=1$.

Comparing different cases it is advisable to set $l*=1$. The resulting mean health state level for the stochastic process is: $H_{mean}=l*/2=(l/\sigma)/2=1/2$. The total health state is estimated as: $H_{total}=(OA)*(OB)/2=l*.(l*/b*)/2=(l*/b*)/2$.

The estimates of Weitz and Fraser for Carey Medfly data by the maximum likelihood estimation method give: $\sigma*=0.0975$ corresponding to $l*=1/\sigma*=10.256$ and $b*=0.448$.

Our estimates by a Levenberg–Marquardt nonlinear regression analysis algorithm provide: $l*=10.532$ corresponding to $\sigma*=1/l*=0.0949$ and $b*=0.480$.

The total health state is $H_{total}=(1/b*)/2=11.446$ for Weitz and Frazer and 10.967 for our estimates.

The advanced method applying (5) for the same data sets provides the following parameter estimates:

$c*=1.321$, $b*=0.2370$ and $l*=8.572$. The total health state is $H_{total}=12.211$.

## 2.2 The Main Points in Estimating the Health State of a Population

1. To introduce a "system of measure" a "metric" that is independent of the related situation
2. To effectively estimate the health state
3. To reproduce the population behavior by means of finding the death probability density, the life expectancy and other measures
4. To compare the health state in various countries and populations for the same or various time periods
5. To find a measure of the health state for all the age period by estimating a measure of the total health state as a summation or integration of all the period of the life span
6. To be able to calculate to cost of the health improvement of the population
7. To estimate the insurance cost and the insurance policies

### 3   The More General Case

The more general case for the health state (5) is of the form

$$H_x = l - (bx)^c \tag{14}$$

As in the previous case and using (8), this form for the health status provides the following probability density function for the first exit time from a barrier set at zero level, that is:

$$g(x) = \frac{|\, l + (c-1)(bx)^c\,|}{\sigma\sqrt{2\pi\, x^3}}\, e^{-\frac{(l-(bx)^c)^2}{2\sigma^2 x}} \tag{15a}$$

By using the transformation

$$b* = b/\sigma^{(1/c)} \text{ and } l* = l/\sigma \tag{15b}$$

we arrive in the following (16) form providing the new parameters $b*$ and $l*$ as functions of $\sigma$.

$$g(x) = \frac{|\, l/\sigma + (c-1)[(b/\sigma^{1/c}\, x)^c]\,|}{\sqrt{2\pi\, x^3}}\, e^{-\frac{(l/\sigma - (b\sigma^{1/c}\, x)^c)^2}{2x}} =$$

$$= \frac{|\, l* + (c-1)(b*\, x)^c\,|}{\sqrt{2\pi\, x^3}}\, e^{-\frac{(l* - (b*\, x)^c)^2}{2x}} \tag{16}$$

The following alternative of (14) accepts the simpler transformation $b* = b/\sigma$ and $l* = l/\sigma$

$$H_x = l - bx^c \tag{17}$$

However, during the applications, the parameter $b$ is very small causing problems in the fitting process.

Both (14) and (17) provide the same measures for the total health state. This is achieved by calculating the integral from age zero ($x=0$) to age $x=T$ of the zero health state.

$$H_{total} = \int_0^T l* - (b*\, x)^c\, dx = \left[ xl* - \frac{(b*)^c\, x^{(c+1)}}{c+1} \right]_0^T = Tl* \frac{c}{c+1} \tag{18}$$

Note that from (14) $H=0$ for $T=l*^{(1/c)}/b*$.

The very important achievement is that by setting $l*=1$ in (18) the total health state is given by:

$$H_{total} = T \frac{c}{c+1} = \frac{l*^{(1/c)}}{b*} \cdot \frac{c}{c+1} \tag{19}$$

Note that the total health state is the more convenient estimator for the comparisons between different populations and countries. It is also in accordance with the main part of the life expectancy at birth ranking though a refined ranking emerges for some countries.

Accordingly the mean health state $H_{mean}$ is given by:

$$H_{mean} = \frac{c}{c+1} \tag{20}$$

We can easily find that for the simple case ($c=1$) the mean health state is $H_{mean}=1/2$ and the total health state is $H_{total}=T/2=(l*/b*)/2$.

Note that the total health state is the more convenient estimator for the comparisons between different populations and countries. The main idea is coming from establishing a health state level equal to unity (see Figure 4A and 4B) at birth by setting $l*=1$. This is an independent measure for every population and country. The next step is to find the total health state by evaluating the integral (18) providing the simple measure in (19). By using the transformation (15b) the right hand side of (19) is given by

$$H_{total} = T \frac{c}{c+1} = \frac{l*^{(1/c)}}{b*} \cdot \frac{c}{c+1} = \frac{l^{(1/c)}/\sigma^{(1/c)}}{b/\sigma^{(1/c)}} \cdot \frac{c}{c+1} = \frac{l^{(1/c)}}{b} \cdot \frac{c}{c+1} \tag{21}$$

The very important achievement from (21) is that the total health state is independent of $\sigma$. When the exponent $c$ is large the total health state is of the order of $l^{(1/c)}/b$. This is a simple method expressing the health state in terms of age as it is done for the life expectancy.

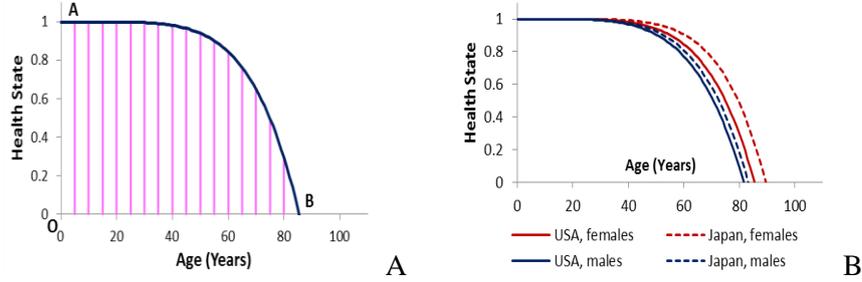

Fig. 4. A. Health State versus age and total health state area and
B. Comparing Health State

## 4    Connection of the health state curve to the survivorship curve

As it was presented earlier the health state curve is independent of $\sigma$. Instead the survivorship curve $l(x)$ or survival curve is connected to $\sigma$ as can be verified by stochastic simulations. Even more the survivorship curve is connected with the death probability which includes $\sigma$ as is expressed in (22).

$$l(x) = 1 - \int_0^x g(s)ds = \int_0^x \frac{|l + (c-1)(bs)^c|}{\sigma\sqrt{2\pi\,s^3}} e^{-\frac{(l-(bs)^c)^2}{2\sigma^2 s}}\,d(s) \quad (22)$$

Where the integral is normalized so that

$$\int_0^\infty g(s)ds = 1 \quad (23)$$

Recalling (8) and with (24) very important relation for Life Expectancy is given by

$$LE = \int_0^x sg(s)ds = \int_0^x \frac{|H_s - sH_s'|}{\sigma\sqrt{2\pi s}} e^{-\frac{H_s^2}{2\sigma^2 s}} \quad (24)$$

From the last formula the relation between the life expectancy and the health state is clarified.

For the simple case for the health state $H_x = l - bx$ the resulting relation is

$$LE = \int_0^x sg(s)ds = \int_0^x \frac{l}{\sigma\sqrt{2\pi s}} e^{-\frac{(l-bs)^2}{2\sigma^2 s}} \quad (25)$$

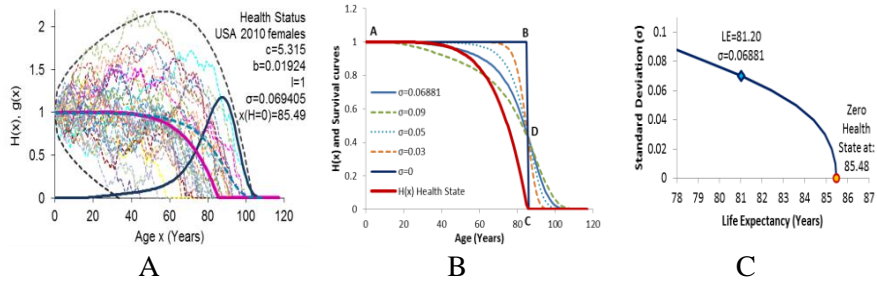

Fig. 5. A. Stochastic Simulations, B. Health State and Survival and C. LE and $\sigma$

Following the above theory we have estimated the health state for the females in USA for 2010 and we have reproduced the results by applying stochastic simulations (see Figure 5A). The health state curve for USA females in 2010 is illustrated by the heavy curve (see Figure 5B). The corresponding survival curve (continuous line) for this related case is presented with $\sigma=0.06881$. The curve with higher value for $\sigma=0.09$ is presented by a dashed line whereas the lower values for $\sigma$ ($\sigma=0.05$, dotted light curve and $\sigma=0.03$ dashed upper curve) appear as well. The very important case with $\sigma=0$ corresponding to the deterministic case is illustrated with a heavy line (ABDC). The latter ends at the point C at the zero health state age. This is the case with the total rectangularization process. In this case the Life Expectancy asymptotically approaches the zero health state age (point C of Figure 5A). All the survival curves cross around the point D at the zero health state age.

## 5 Comparing the Health State or Health Status and Life Expectancy

From equation (22) for the survivorship we have estimated the Life Expectancy for various values of the standard deviation $\sigma$. The results are illustrated in Figure 5C. The Life Expectancy at Birth is systematically related to the standard deviation $\sigma$. At $\sigma=0$ the Life Expectancy is equal to the age at zero Health Status (application for USA, females, 2010). The life expectancy as it is given in the life table from Human Mortality

Database is 81.20 years and we estimated that this is achieved for $\sigma$=0.06881.

### 5.1    The time development of the standard deviation $\sigma$

As far as the standard deviation $\sigma$ is estimated by the proposed method we have calculated the related figures for USA males and females and presented in Figure 6. The standard deviation and thus the dispersion are larger for males especially for the years after 1990.

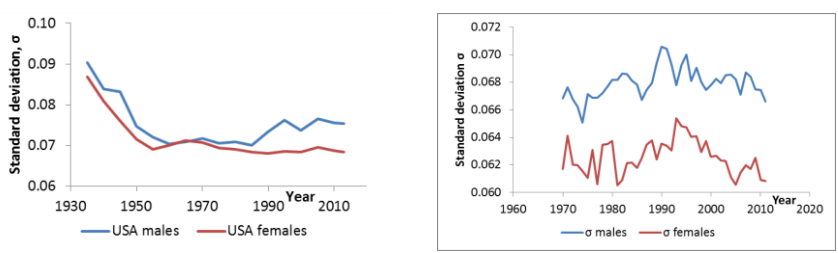

Fig. 6. A. The time development of the standard deviation $\sigma$ for USA and B. for Switzerland

## 6. Health State, Life Expectancy and HALE estimates

The proposed methodology for estimating the Health State or Health Status of a population provides a tool for comparing the HS with the healthy life expectancy as it is presented by the HALE measures of the World Health Organization (WHO). Related information and publications are included in very may references (see Murray et al. 2015, Robine 1999, Salomon 2012 and WHO annexes from 2000-2015 and many publications in Lancet).

The very important property of the HS measure is that it is independent of the dispersion expressed by the standard deviation $\sigma$. It will be the same if the dispersion is large or small. Instead the life expectancy is influenced by the dispersion as it was provided in formulas (22) and (24) and presented in Figure 5B. So far we can estimate the LE and the HS for every time period as far as life table data are provided.

Table I includes our estimates for the HS and LE for 36 countries. These countries are included in the Human Mortality Database and are extensively studied. The countries are ranked according to the HS.

**TABLE I**

| Males | | | | Females | | |
|---|---|---|---|---|---|---|
| **Country** | **HS** | **LE** | | **Country** | **HS** | **LE** |
| Switzerland | 70.58 | 80.05 | | Japan | 76.49 | 86.30 |
| Australia | 70.57 | 79.87 | | France | 75.87 | 84.69 |
| New Zealand | 69.93 | 79.16 | | Switzerland | 75.46 | 84.39 |
| France | 69.67 | 78.04 | | Spain | 74.78 | 85.01 |
| Canada | 69.56 | 79.17 | | Australia | 74.74 | 84.25 |
| Sweden | 69.51 | 79.51 | | Luxembourg | 74.30 | 83.18 |
| Japan | 69.51 | 79.56 | | Finland | 74.23 | 83.24 |
| Italy | 69.30 | 79.49 | | Italy | 74.21 | 84.49 |
| Spain | 68.97 | 79.01 | | Canada | 73.95 | 83.51 |
| Norway | 68.96 | 78.84 | | Sweden | 73.95 | 83.47 |
| Austria | 68.42 | 77.67 | | Austria | 73.89 | 83.13 |
| Luxembourg | 68.40 | 77.94 | | New Zealand | 73.85 | 82.96 |
| UK | 68.31 | 78.37 | | Belgium | 73.78 | 82.65 |
| Ireland | 68.24 | 78.26 | | Norway | 73.69 | 83.15 |
| Greece | 68.16 | 77.99 | | Netherlands | 73.42 | 82.72 |
| Portugal | 67.96 | 76.74 | | Germany | 73.30 | 82.71 |
| Belgium | 67.94 | 77.38 | | Slovenia | 72.61 | 82.62 |
| Israel | 67.92 | 79.70 | | Portugal | 72.31 | 83.04 |
| Netherlands | 67.91 | 78.78 | | Ireland | 72.29 | 82.77 |
| Finland | 67.60 | 76.72 | | UK | 72.17 | 82.35 |
| USA | 67.52 | 76.38 | | Israel | 72.06 | 83.50 |
| Germany | 67.42 | 77.67 | | USA | 71.86 | 81.22 |
| Denmark | 66.60 | 77.12 | | Greece | 71.71 | 83.17 |
| Scotland | 66.11 | 76.24 | | Taiwan | 71.26 | 82.37 |
| Taiwan | 65.72 | 76.24 | | Poland | 70.99 | 80.47 |
| Slovenia | 65.70 | 76.28 | | Estonia | 70.93 | 80.55 |
| Czech Republic | 63.24 | 74.47 | | Czech Republic | 70.90 | 80.64 |
| Poland | 60.35 | 72.16 | | Lithuania | 70.54 | 78.75 |
| Slovakia | 59.69 | 71.73 | | Denmark | 70.13 | 81.33 |
| Bulgaria | 59.47 | 70.31 | | Scotland | 70.09 | 80.63 |
| Estonia | 59.44 | 70.83 | | Slovakia | 68.93 | 79.15 |
| Lithuania | 55.15 | 67.55 | | Bulgaria | 68.82 | 77.25 |
| Ukraine | 54.61 | 65.20 | | Belarus | 67.16 | 76.49 |
| Latvia | 53.55 | 67.43 | | Latvia | 66.95 | 77.42 |
| Russia | 52.68 | 63.06 | | Russia | 66.81 | 74.87 |
| Belarus | 52.06 | 64.60 | | Ukraine | 66.27 | 75.19 |

There is relatively good agreement with the LE rankings although differences appear as it was expected. The measure based on HS is also a strong methodological tool avoiding dispersion. We can call it an **absolute measure of the Health State (HE)** of a population.

Following our introduction of the HS as an absolute measure of health state we search if this estimate could be a measure or metric of the Healthy Life Expectancy estimates as are expressed by the so-called HALE system. However, HALE estimates are strongly influenced by the dispersion $\sigma$ and could not be compared with the HS. The various healthy life expectancy (HLE) estimates by using only the life table data are presented in Skiadas (ArXiv.org, March 2016) with good results compared to the HALE estimates. Here we use one of these estimates called Direct Estimation from the Life Tables (see Skiadas, ArXiv.org, March 2016, pp. 7-8). This is illustrated in Figures for few of the countries presented in Table I. Fortunately the HALE data provided include the estimated confidence intervals thus improving comparisons as is presented in Figures 7A-7N where the confidence bars appear. The HS is expressed by the heavy blue curve and the HLE by light red curve whereas the HALE estimates are presented rhombus with confidence bars.

It is demonstrated is that the Health State is growing for the years studied (more than 40 years starting 1970). Only for males in the Czech Republic was a decline few years before 1990 and then a continuous increase until nowadays. The healthy life expectancy (HLE) was not increasing as fast as the Health State (HS) in several cases.

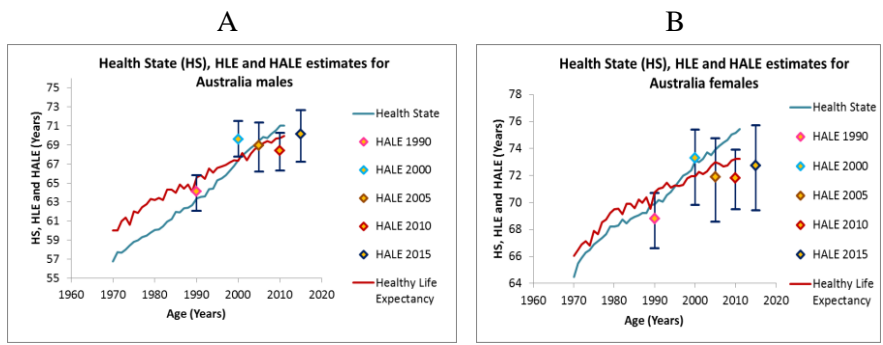

C

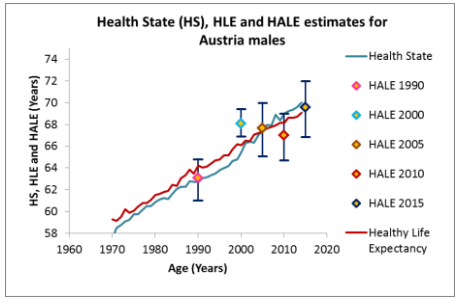

D

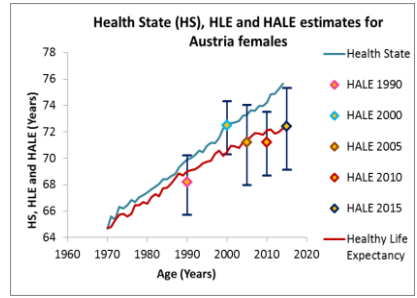

E

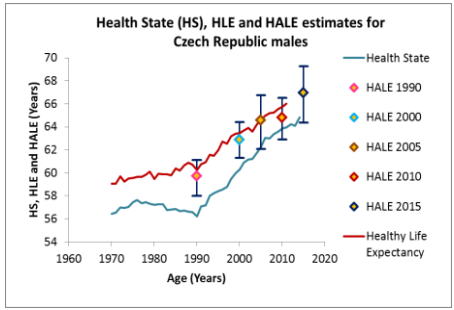

F

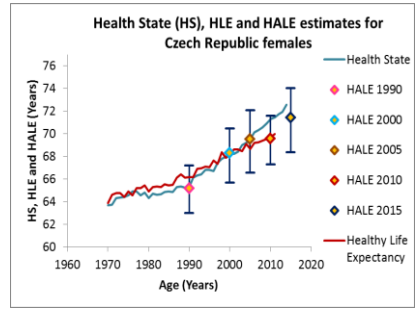

G

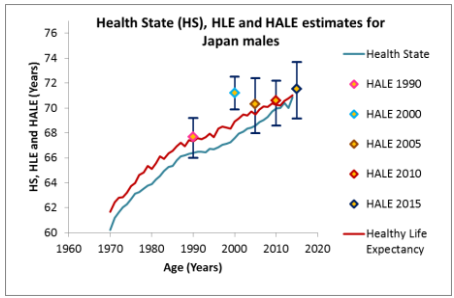

H

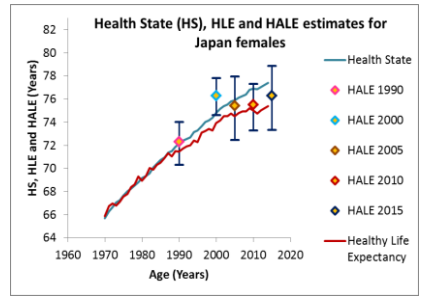

I

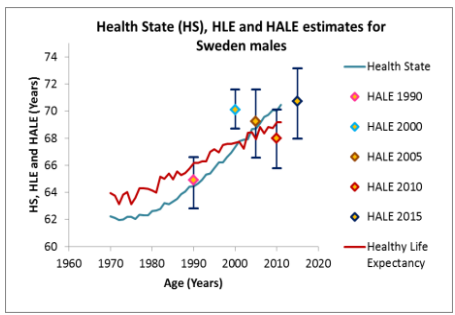

J

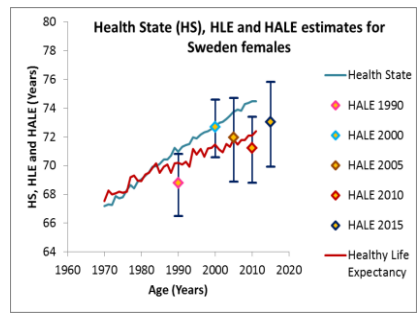

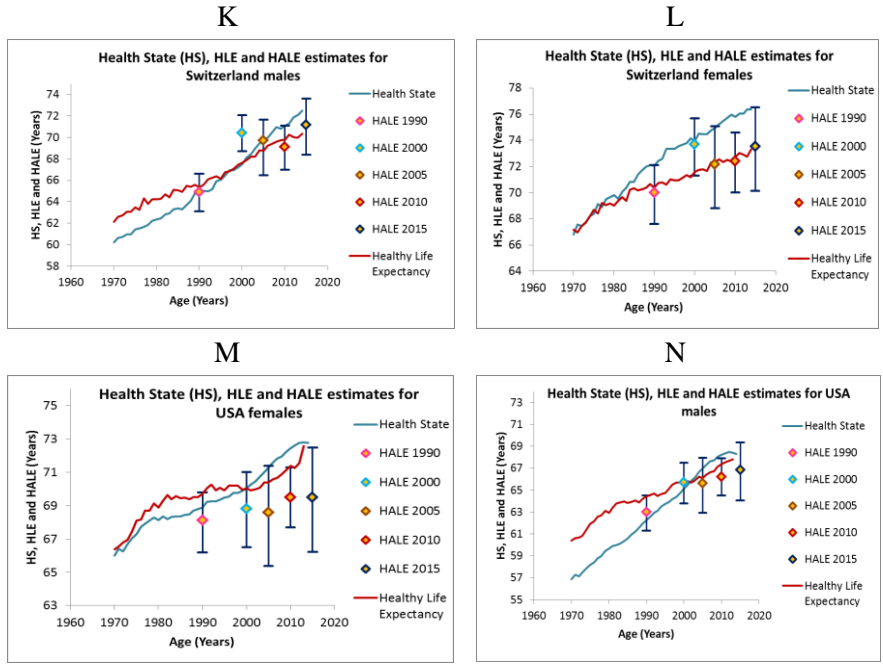

Fig. 7A-7N. Health State, Healthy Life Expectancy and HALE

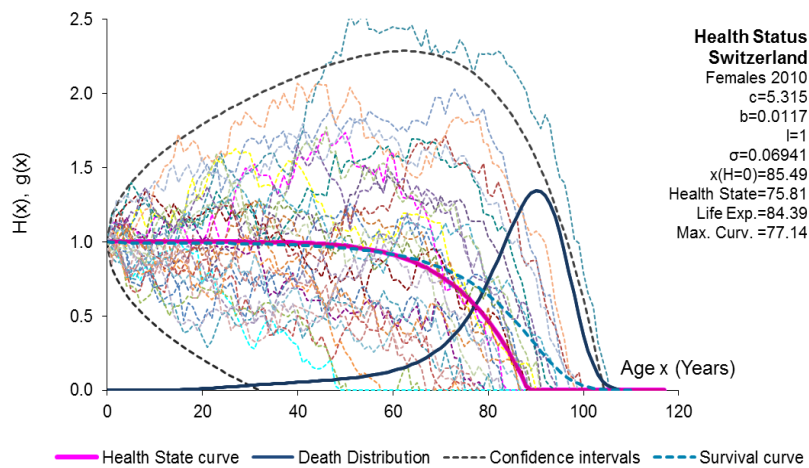

Fig. 8. Health state curve, survival curve, death distribution, stochastic paths and confidence intervals for females in Switzerland 2010.

## 7 Further Discussion

We have introduced and estimated the Health State of a population along with the associated Health State Curve and related parameters as the Age of zero Health State and the Age of the Maximum Curvature of the health state curve. This is also an important age year where the maximum decline of the health state appears. Furthermore we have connected the health state with the survival curve and the standard deviation $\sigma$ as it is estimated from the Death Probability Density Function. Stochastic simulations are also important to verify the theory and to reproduce the provided data after a quite large number of stochastic paths generated. The main part of the theoretical and applied findings is illustrated in Figure 8 for females in Switzerland 2010. The model parameters estimated are b=0.0117, c=5.315, $\sigma$=0.06941 for l=1. The Health State estimated (HS=78.85) is precisely the area under the Health State Curve whereas the Life Expectancy estimated (LE=84.39) is the area under the Survival Curve. The Maximum Curvature is at 77.14 years defining the maximum deterioration stage.

## 8 Conclusion

We have presented and applied the general theory of the health state of a population in connection to the survival curves and life expectancy. We have provided a measure for the Health State or Health Status which is independant of the dispersion parameter expressed by the standard deviation $\sigma$. We also have provided enough evidence of the relation of life expectancy and the health state or health status estimator. We complete our work with stochastic simulations thus effectively reproducing the death distribution. We also provide illustrations with our estimated Health Status and a comparative approach of Healthy Life Expectancy and the HALE measures in several countries.